\documentclass[11pt]{article} 
\usepackage{amsmath}
\usepackage{amsfonts,amssymb}

\begin{document}

\newcommand{\nl}{\nonumber\\}
\newcommand{\nnl}{\nl[6mm]}
\newcommand{\nle}{\nl[-2.5mm]\\[-2.5mm]}
\newcommand{\nlb}[1]{\nl[-2.0mm]\label{#1}\\[-2.0mm]}
\newcommand{\ab}{\allowbreak}

\renewcommand{\leq}{\leqslant}
\renewcommand{\geq}{\geqslant}

\renewcommand{\theequation}{\thesection.\arabic{equation}}
\let\ssection=\section
\renewcommand{\section}{\setcounter{equation}{0}\ssection}

\newcommand{\be}{\bes}
\newcommand{\ee}{\ees}
\newcommand{\bes}{\begin{eqnarray}}
\newcommand{\ees}{\end{eqnarray}}
\newcommand{\eens}{\nonumber\end{eqnarray}}
\newcommand{\barr}{\begin{array}}
\newcommand{\earr}{\end{array}}

\renewcommand{\/}{\over}
\renewcommand{\d}{\partial}
\newcommand{\Dslash}{\hbox{$D\kern-2.4mm/\,$}}
\newcommand{\dd}[1]{\ab\delta/\delta {#1}}

\newcommand{\bra}[1]{\big{\langle}#1\big{|}}
\newcommand{\ket}[1]{\big{|}#1\big{\rangle}}

\newcommand{\summ}[1]{\sum_{|\mm|\leq #1}}
\newcommand{\sumnm}{\sum_{|\nn|\leq |\mm|}}
\newcommand{\summn}[1]{\sum_{|\nn|\leq |\mm|\leq #1}}

\newcommand{\mm}{{\mathbf m}}
\newcommand{\nn}{{\mathbf n}}
\newcommand{\phim}{\phi_{,\mm}}
\newcommand{\phin}{\phi_{,\nn}}
\newcommand{\pim}{{\pi}^{,\mm}}
\newcommand{\pin}{{\pi}^{,\nn}}

\newcommand{\fa}{\phi_\alpha}
\newcommand{\fb}{\phi_\beta}
\newcommand{\pa}{\pi^\alpha}
\newcommand{\pb}{\pi^\beta}

\renewcommand{\fam}{\phi_{\alpha,\mm}}
\newcommand{\fan}{\phi_{\alpha,\nn}}
\newcommand{\pam}{\pi^{\alpha,\mm}}
\newcommand{\pan}{\pi^{\alpha,\nn}}

\newcommand{\fbm}{\phi_{\beta,\mm}}
\newcommand{\fbn}{\phi_{\beta,\nn}}
\newcommand{\pbm}{\pi^{\beta,\mm}}
\newcommand{\pbn}{\pi^{\beta,\nn}}

\newcommand{\Ea}{\EE^\alpha}
\newcommand{\Eb}{\EE^\beta}

\newcommand{\fs}{\phi^*}
\newcommand{\fsa}{\phi^{*\alpha}}
\newcommand{\fsb}{\phi^{*\beta}}
\newcommand{\psa}{\pi^*_\alpha}
\newcommand{\psb}{\pi^*_\beta}

\newcommand{\wfa}{{\overline\phi}_\alpha}
\newcommand{\wpa}{{\overline\pi}{}^\alpha}
\newcommand{\wfsa}{{\overline\phi}{}^{*\alpha}}
\newcommand{\wpsa}{{\overline\pi}{}^*_\alpha}
\newcommand{\wfam}{{\overline\phi}_{\alpha,\mm}}
\newcommand{\wfbn}{{\overline\phi}_{\beta,\nn}}
\newcommand{\wpam}{{\overline\pi}{}^{\alpha,\mm}}
\newcommand{\wfsam}{{\overline\phi}{}^{*\alpha}_{,\mm}}
\newcommand{\wfsan}{{\overline\phi}{}^{*\alpha}_{,\nn}}
\newcommand{\wfsbn}{{\overline\phi}{}^{*\beta}_{,\nn}}
\newcommand{\wpsam}{{\overline\pi}{}^{*,\mm}_\alpha}
\newcommand{\wpsan}{{\overline\pi}{}^{*,\nn}_\alpha}
\newcommand{\wpsbn}{{\overline\pi}{}^{*,\nn}_\beta}
\newcommand{\wpsbm}{{\overline\pi}{}^{*,\mm}_\beta}
\newcommand{\wba}{{\overline b}{}^a}
\newcommand{\wca}{{\overline c}_a}
\newcommand{\wbam}{{\overline b}{}^a_{,\mm}}
\newcommand{\wban}{{\overline b}{}^a_{,\nn}}
\newcommand{\wbbn}{{\overline b}{}^b_{,\nn}}
\newcommand{\wcam}{{\overline c}{}^{,\mm}_a}
\newcommand{\wcbm}{{\overline c}{}^{,\mm}_b}

\newcommand{\fsam}{\phi^{*\alpha}_{,\mm}}
\newcommand{\fsan}{\phi^{*\alpha}_{,\nn}}
\newcommand{\fsbn}{\phi^{*\beta}_{,\nn}}
\newcommand{\psam}{\pi^{*,\mm}_\alpha}
\newcommand{\psbm}{\pi^{*,\mm}_\beta}
\newcommand{\psbn}{\pi^{*,\nn}_\beta}
\newcommand{\Eam}{\EE^\alpha_{,\mm}}
\newcommand{\bam}{b^a_{,\mm}}
\newcommand{\ban}{b^a_{,\nn}}
\newcommand{\bbn}{b^b_{,\nn}}
\newcommand{\cam}{c^{,\mm}_a}
\newcommand{\cbm}{c^{,\mm}_b}
\newcommand{\ram}{r^a_{,\mm}}

\newcommand{\ord}{o}
\newcommand{\ordg}{\varsigma}

\newcommand{\repi}{\rep^{(i)}}
\newcommand{\Mi}{M^{(i)}}
\newcommand{\mri}{(-)^i {r\choose i}}

\newcommand{\si}{\sigma}
\newcommand{\eps}{\epsilon}
\newcommand{\dlt}{\delta}
\newcommand{\om}{\omega}
\newcommand{\al}{\alpha}
\newcommand{\bt}{\beta}
\newcommand{\ka}{\kappa}
\newcommand{\la}{\lambda}
\newcommand{\vth}{\vartheta}
\renewcommand{\th}{\theta}
\newcommand{\rep}{\varrho}

\newcommand{\xmu}{\xi^\mu}
\newcommand{\ynu}{\eta^\nu}
\newcommand{\dmu}{\d_\mu}
\newcommand{\dnu}{\d_\nu}
\newcommand{\drho}{\d_\rho}

\newcommand{\ssu}{su(3)\!\oplus\! su(2)\!\oplus\! u(1)}

\newcommand{\no}[1]{{\,:\kern-0.7mm #1\kern-1.2mm:\,}}

\newcommand{\vect}{{\mathfrak{vect}}}
\newcommand{\svect}{{\mathfrak{svect}}}
\newcommand{\map}{{\mathfrak{map}}}
\newcommand{\mb}{{\mathfrak{mb}}}
\newcommand{\vle}{{\mathfrak{vle}}}

\renewcommand{\L}{{\cal L}}
\newcommand{\J}{{\cal J}}
\newcommand{\Lxi}{\L_\xi}

\newcommand{\im}{{\rm im}}
\newcommand{\e}{{\rm e}}
\renewcommand{\div}{{\rm div}}
\newcommand{\afn}{{\rm afn\,}}
\newcommand{\til}{{\tilde{\ }}}

\newcommand{\Np}[1]{{N+p\choose N #1}}
\newcommand{\Npr}{{N+p-r\choose N-r}}
\newcommand{\ritwo}{(-)^i {r-2\choose i-2}}
\newcommand{\rione}{(-)^i {r-1\choose i-1}}

\newcommand{\qmu}{q^\mu}
\newcommand{\qnu}{q^\nu}
\newcommand{\qrho}{q^\rho}
\newcommand{\pmu}{p_\mu}
\newcommand{\pnu}{p_\nu}

\newcommand{\dNx}{d^N\!x}
\newcommand{\dxdt}{\dNx\,dt}
\renewcommand{\gg}{\oj\oplus gl(N)}
\newcommand{\dmap}{\vect(N)\ltimes \map(N,\oj)}

\newcommand{\larroww}[1]{{\ \stackrel{#1}{\longleftarrow}\ }}
\newcommand{\brep}[1]{{\bf{\underline{#1}}}}

\newcommand{\tr}[1]{{\rm tr}_{#1}\kern0.7mm}
\newcommand{\oj}{{\mathfrak g}}
\newcommand{\hh}{{\mathfrak h}}

\newcommand{\U}{{\cal U}}
\newcommand{\QQ}{{\cal Q}}
\newcommand{\PP}{{\cal P}}
\newcommand{\EE}{{\cal E}}
\newcommand{\FF}{{\cal F}}
\newcommand{\II}{{\cal I}}
                                            
\newcommand{\TT}{{\mathbb T}}
\newcommand{\RR}{{\mathbb R}}
\newcommand{\CC}{{\mathbb C}}
\newcommand{\ZZ}{{\mathbb Z}}
\newcommand{\NN}{{\mathbb N}}

\title{{Multi-dimensional Virasoro algebra and quantum gravity}}

\author{T. A. Larsson \\
Vanadisv\"agen 29, S-113 23 Stockholm, Sweden\\
email: thomas.larsson@hdd.se}

\maketitle 
\begin{abstract} 
I review the multi-dimensional generalizations of the Virasoro algebra,
i.e. the non-central Lie algebra extensions of the algebra
$\vect(N)$ of general vector fields in $N$ dimensions, and its Fock
representations. Being the Noether symmetry of background independent
theories such as $N$-dimensional general relativity, this algebra is
expected to be relevant to the quantization of gravity. To this end,
more complicated modules which depend on dynamics in the form of
Euler-Lagrange equations are described. These modules can apparently
only be interpreted as quantum fields if spacetime has four dimensions
and both bosons and fermions are present.
\end{abstract}

\medskip
\noindent In: Mathematical physics research at the leading edge  \hfill\break
ed: Charles V. Benton, pp 91-111		       \hfill\break
2004 Nova Science Publishers, Inc.		       \hfill\break
ISBN 1-59033-905-3				       \hfill

\newpage

\section{Introduction} 

It is widely recognized that a candidate theory of quantum gravity must 
be general-covariant, i.e. it must carry a representation of
the full spacetime diffeomorphism group \cite{GR99}. Since tensor
densities are the classical modules of this group \cite{Rud74,VGG75},
this implies in particular that tensor calculus is the appropriate
language of general relativity. However, experience with conformal field
theory teaches us that the physically interesting representations are
projective, i.e. that the corresponding Lie algebra acquires an
extension. It is thus natural to look for a generalization of the
Virasoro algebra to $N$ dimensions, i.e. an extension $Vir(N)$ of the
algebra $\vect(N)$ of general vector fields (on the
$N$-dimensional torus, say). Other names for $\vect(N)$ are
diffeomorphism algebra or generalized Witt algebra; algebraists often
denote it by $W_N$ in honor of Witt.

The first interesting representations of $Vir(N)$ were constructed by Rao
and Moody \cite{RM94}, and the Fock/vertex operator modules were
essentially understood in \cite{BBS01,Bil02,Lar98}. However, general
covariance by itself is certainly not enough to describe gravity;
information about the Einstein equations must somehow be introduced. In
\cite{Lar02} a class of modules with this property was described. To each
dynamical system, we can associate a family of representations of its
Noether symmetry algebra; this is quantization in the sense that the
brackets acquire non-trivial quantum corrections. In particular, chosing
the dynamical system to be general relativity gives us a kind of quantum
gravity, which is well-defined as a $Vir(N)$ module.

A crucial step in the construction of Fock modules is the replacement of
all fields by $p$-jets, where $p$ is a finite integer. In order to have a
field theory interpretation, it must be possible to
take the limit $p\to\infty$. This limit is problematic, because the
abelian charges (the higher-dimensional analogues of the central charge)
diverge, but the leading divergences can be cancelled by a clever choice
of field content. With some natural assumptions of the form on the
Euler-Lagrange equations (second order for bosons and first order for
fermions, and Noether identities of one order higher), it turns out that
the finiteness conditions can only be satisfied if spacetime has four
dimensions and there are two bosons for every three fermions with the
na\"\i ve counting of degrees of freedom; this relation holds in the
standard model coupled to gravity. Note that already the prediction of
both bosons and fermions is quite remarkable without superalgebras.
However, the same argument also requires new gauge symmetries, including
fermionic ones, perhaps indicating the need for some new physics.

The reason why $Vir(N)$ went unnoticed for several decades is that
it is not a {\em central} extension. The fact that $\vect(N)$ has no
central extension when $N>1$ has been rediscovered many times
\cite{Dzhu84,RS76,RSS89}; see \cite{GLS97} for the classification
of central extensions of simple Lie superalgebras of vector fields.
However, it does have two inequivalent abelian but non-central 
extensions which both reduce to the Virasoro algebra when $N=1$
\cite{Lar91,RM94}.

Dzhumadildaev has classified all extensions of $\vect(N)$ by
modules of tensor densities \cite{Dzhu96,Lar00}. In contrast, the two
Virasoro-like cocycles involve modules of closed $(N-1)$-forms, which
are not tensor modules but rather submodules thereof. However, the two
Virasoro cocycles are closely related to his cocycles $\psi^W_3$ and
$\psi^W_4$. It can be noted that most (possibly all) extensions by
tensor modules are limiting cases of trivial extensions, in the sense
that one can construct a one-parameter family of trivial cocycles
reducing to the non-trivial cocycle for a critical value of the
parameter. In contrast, the Virasoro-like cocycles are not limits of
trivial cocycles, because modules of closed forms do not depend on any
continuous parameters. They also arise naturally in Fock
representations, and having a representation
theory is of course essential for any application to physics.

\section{Multi-dimensional Virasoro algebra}
\label{sec:VirN}

To make the connection to the Virasoro algebra very explicit, it is
instructive to write down the brackets in a Fourier basis. Start with
the Virasoro algebra $Vir$:
\be
[L_m, L_n] = (n-m)L_{m+n} - {c\/12} (m^3-m) \delta_{m+n},
\ee
where $\delta_m$ is the Kronecker delta. When $c=0$, 
$L_m = -i \exp(imx) d/dx$, $m \in \ZZ$. 
The element $c$ is central, meaning that it commutes with all of $Vir$;
by Schur's lemma, it can therefore be considered as a c-number.
Now rewrite $Vir$ as
\bes
[L_m, L_n] &=& (n-m)L_{m+n} + c m^2 n S_{m+n}, \nl
{[}L_m, S_n] &=& (n+m)S_{m+n}, \nle
{[}S_m, S_n] &=& 0, \nl
m S_m &=& 0.
\eens
It is easy to see that the two formulations of $Vir$ are equivalent
(I have absorbed the linear cocycle into a redefinition of $L_0$).
The second formulation immediately generalizes to $N$ dimensions.
The generators are $L_\mu(m) = -i \exp(i m_\rho x^\rho) \dmu$ and
$S^\mu(m)$, where $x = (x^\mu)$, $\mu = 1, 2, ..., N$ is a point in 
$N$-dimensional space and $m = (m_\mu)$. The Einstein convention is used;
repeated indices, one up and one down, are implicitly summed over. 
The defining relations are
\bes
[L_\mu(m), L_\nu(n)] &=& n_\mu L_\nu(m+n) - m_\nu L_\mu(m+n) \nl 
&&  + (c_1 m_\nu n_\mu + c_2 m_\mu n_\nu) m_\rho S^\rho(m+n), \nl
{[}L_\mu(m), S^\nu(n)] &=& n_\mu S^\nu(m+n)
 + \delta^\nu_\mu m_\rho S^\rho(m+n), \nle
{[}S^\mu(m), S^\nu(n)] &=& 0, \nl
m_\mu S^\mu(m) &=& 0.
\eens
This is an extension of $\vect(N)$ by the abelian ideal with basis 
$S^\mu(m)$. This algebra is even valid globally on the 
$N$-dimensional torus $\TT^N$.
Geometrically, we can think of $L_\mu(m)$ as a vector field 
and $S^\mu(m) = \eps^{\mu\nu_2..\nu_N}S_{\nu_2..\nu_N}(m)$ 
as a dual one-form (and $S_{\nu_2..\nu_N}(m)$ as an $(N-1)$-form);
the last condition expresses closedness. 

The cocycle proportional to $c_1$ was discovered by 
Rao and Moody \cite{RM94}, and the one proportional to $c_2$ by
myself \cite{Lar91}. There is also a similar multi-dimensional 
generalization of affine Kac-Moody algebras,
presumably first written down by Kassel \cite{Kas85}.
The multi-dimensional Virasoro and 
affine algebras are often refered to as ``Toroidal Lie algebras''
in the mathematics literature
\cite{BB98,BBS01,Bil97,Bil02,Bil03,MRY90,RMY92,Rao02}.

\section{ Failure of the na\"\i ve approach to Fock representations }
\label{sec:rep}

To construct Fock representations of the ordinary Virasoro algebra is
straightforward:
\begin{itemize}
\item
Start from classical modules, i.e. primary fields = scalar densities.
\item
Introduce canonical momenta.
\item
Normal order.
\end{itemize}
However, this approach does simply not work in several dimensions,
because there are problems with normal ordering:
\begin{itemize}
\item
It requires that at least a partial order has been introduced, which 
runs against the idea of diffeomorphism invariance.
\item
Normal ordering of bilinear expressions always results in a {\em central}
extension, but the Virasoro cocycle is non-central when $N\geq 2$.
\item
It is ill defined. Formally, attempts to normal order result in an
{\em infinite} central extension, which of course makes no sense.
\end{itemize}
This problem is the reason why the first interesting representations
of the multi-dimensional Virasoro algebra only appeared a quarter 
century after their one-dimensional siblings. That it is a real problem 
can be seen by looking at the early (and failed) attempts in
\cite{FR91,Lar89,Lar91,RSS89}. It was only with the breakthrough in
\cite{RM94} that progress became possible. This was followed by a
number of papers by different authors 
\cite{BB98,BBS01,Bil97,Lar97a,Lar98,Lar01a},
where the Rao-Moody construction was generalized in several ways and 
the underlying geometry explained.

The main idea, as described in the physics-flavored language of 
\cite{Lar98}, is as follows:
\begin{itemize}
\item
The arena is not just $N$-dimensional spacetime, but spacetime with a
marked one-dimensional curve on it, the observer's trajectory.
\item
All fields must be expanded in a Taylor series around the points of
the observer's trajectory, truncated at some arbitrary but fixed 
order $p$. In other words, we pass from the fields to the corresponding
$p$-jets, or rather trajectories in jet space.
\item
We now have a classical realization on finitely many functions of a
single variable (the parameter along the trajectory), which is
precisely the situation where normal ordering applies. 
\item
Introduce canonical momenta to the jets (not to the fields) and normal
order with respect to frequency. This yields a realization of $Vir(N)$.
Since the classical realization on jets is non-linear, the extension
is non-central.
\item
The classical realization is highly reducible, since each point on
the trajectory transforms independently of its neighbors. To lift this 
degeneracy, we introduce an additional $\vect(1)$ factor, describing 
reparametrizations of the observer's trajectory.
The relevant algebra thus becomes the 
extension of $\vect(N)\oplus\vect(1)$ by its four Virasoro-like cocycles.
\item
The reparametrization symmetry can be eliminated with a constraint, but
then one of the spacetime direction (``time'') is singled out
\cite{Lar98}. Two of the four Virasoro-like cocycles of $DRO(N)$ 
transmute into the complicated
anisotropic cocycles found in \cite{Lar97a}; these are colloquially known
as the ``messy cocycles''. By further specialization to scalar-valued
zero-jets on the torus, the results of Rao and Moody are recovered.
\end{itemize}

The Kassel extension of current algebras can be treated along similar
lines.

\section{DGRO algebra}

By the arguments in the previous section one is lead to study the DGRO 
{\em (Diffeomorphism, Gauge, Repara\-metri\-zation, Observer)} algebra
$DGRO(N,\oj)$, whose ingredients are spacetime diffeomorphisms which 
generate $\vect(N)$,
repara\-metri\-zations of the observer's trajectory which form an
additional $\vect(1)$ algebra, and gauge transformations which generate
a current algebra. Classically, the algebra is $\dmap\oplus\vect(1)$.

Let $\xi=\xmu(x)\dmu$, $x\in\RR^N$, $\dmu = \d/\d x^\mu$,
be a vector field, with commutator 
$[\xi,\eta] \equiv \xmu\dmu\ynu\dnu - \ynu\dnu\xmu\dmu$,
and greek indices $\mu,\nu = 1,2,..,N$ label the 
spacetime coordinates. 
The Lie derivatives $\Lxi$ are the infinitesimal diffeomorphisms, i.e.
the generators of $\vect(N)$.

Let $f = f(t)d/dt$, $t\in S^1$, be a vector field in one dimension. 
The commutator reads 
$[f,g] = (f\dot g - g\dot f)d/dt$, where the dot denotes the $t$ 
derivative: $\dot f \equiv df/dt$. We will also use $\d_t = \d/\d t$
for the partial $t$ derivative.
The choice that $t$ lies on the circle is physically unnatural and is
made for technical simplicity only (quantities can be expanded in
Fourier series). However, this seems to be a
minor problem at the present level of understanding.
Denote the repara\-metri\-zation generators $L_f$.

Let $\map(N,\oj)$ be the current algebra corresponding to the
finite-dimen\-sional semisimple Lie algebra $\oj$ with basis $J^a$, 
structure constants $f^{ab}{}_c$, and Killing metric $\dlt^{ab}$. 
The brackets in $\oj$ are 
\be
[J^a, J^b] = if^{ab}{}_c J^c.
\label{oj}
\ee
A basis for $\map(N,\oj)$ is given by $\oj$-valued functions $X=X_a(x)J^a$ 
with commutator $[X,Y]=if^{ab}{}_c X_aY_bJ^c$. The intertwining
$\vect(N)$ action is given by $\xi X = \xmu\dmu X_a J^a$. Denote the
$\map(N,\oj)$ generators by $\J_X$.

Finally, let $Obs(N)$ be the space of local functionals of the
observer's tractory $\qmu(t)$, i.e. polynomial functions of  
$\qmu(t)$, $\dot q^\mu(t)$, ... $d^k \qmu(t)/dt^k$,  $k$ finite, 
regarded as a commutative algebra. $Obs(N)$ is a $\vect(N)$ module in a 
natural manner.

$DGRO(N,\oj)$ is an abelian but non-central Lie algebra extension of 
$\vect(N) \ltimes \map(N,\oj) \oplus \vect(1)$ by $Obs(N)$:
\[
0 \longrightarrow Obs(N) \longrightarrow DGRO(N,\oj) \longrightarrow
 \vect(N)\ltimes \map(N,\oj)\oplus \vect(1) \longrightarrow 0.
\]
The brackets are given by
\bes
[\Lxi,\L_\eta] &=& \L_{[\xi,\eta]} 
 + {1\/2\pi i}\int dt\ \dot q^\rho(t) 
 \Big\{ c_1 \drho\dnu\xmu(q(t))\dmu\ynu(q(t)) +\nl
&&\quad+ c_2 \drho\dmu\xmu(q(t))\dnu\ynu(q(t)) \Big\}, \nl
{[}\Lxi, \J_X] &=& \J_{\xi X}, \nl
{[}\J_X, \J_Y] &=& \J_{[X,Y]} - {c_5\/2\pi i}\dlt^{ab}
 \int dt\ \dot q^\rho(t)\drho X_a(q(t))Y_b(q(t)), \nl
{[}L_f, \Lxi] &=& {c_3\/4\pi i} \int dt\ 
 (\ddot f(t) - i\dot f(t))\dmu\xmu(q(t)), 
\label{DGRO}\\
{[}L_f,\J_X] &=& 0, \nl
{[}L_f,L_g] &=& L_{[f,g]} 
 + {c_4\/24\pi i}\int dt (\ddot f(t) \dot g(t) - \dot f(t) g(t)), \nl
{[}\Lxi,\qmu(t)] &=& \xmu(q(t)), \nl
{[}L_f,\qmu(t)] &=& -f(t)\dot q^\mu(t), \nl
{[}\J_X, \qmu(t)] &=& {[}\qmu(s), \qnu(t)] = 0,
\eens
extended to all of $Obs(N)$ by Leibniz' rule and linearity.
The numbers $c_1-c_5$ are called {\em abelian charges}, in analogy with
the central charge of the Virasoro algebra. In the
references slightly more complicated extensions are considered, which
depend on three additional abelian charges $c_6-c_8$. However, these
vanish automatically when $\oj$ is semisimple.

\section{ Trajectories in jet space }
\label{sec:Fock}

The classical representations of the DGRO algebra are tensor fields
over $\RR^N \times S^1$ valued in $\oj$ modules. The basis of a 
classical DGRO module $\QQ$ is thus a field 
$\fa(x,t)$, $x\in\RR^N$, $t\in S^1$, 
where $\al$ is a collection of all kinds of indices.
The $DGRO(N,\oj)$ action on $\QQ$ can be succinctly summarized as
\bes
[\Lxi, \fa(x,t)] &=& -\xmu(x)\dmu\fa(x,t)
- \dnu\xmu(x)T^{\bt\nu}_{\al\mu}\fb(x,t), \nl
{[}\J_X, \fa(x,t)] &=& -X_a(x)J^{\bt a}_\al\fb(x,t), 
\\
{[}L_f, \fa(x,t)] &=& -f(t)\d_t\fa(x,t) 
- \la(\dot f(t)-if(t))\fa(x,t).
\eens
Here $J^a = (J^{\al a}_\bt)$ and $T^\mu_\nu = (T^{\al\mu}_{\bt\nu})$ are
matrices satisfying $\oj$ (\ref{oj}) and $gl(N)$, respectively:
\be
[T^\mu_\nu,T^\si_\tau] = 
\dlt^\si_\nu T^\mu_\tau - \dlt^\mu_\tau T^\si_\mu.
\ee

The crucial idea in \cite{Lar98} is to expand all fields in a Taylor 
series around the observer's trajectory and truncate at order $p$, 
before introducing canonical momenta. Hence e.g.,
\be
\fa(x,t) = \summ p {1\/\mm!} \fam(t)(x-q(t))^\mm,
\label{Taylor}
\ee
where $\mm = (m_1, \ab m_2, \ab ..., \ab m_N)$, all $m_\mu\geq0$, is a 
multi-index of length $|\mm| = \sum_{\mu=1}^N m_\mu$,
$\mm! = m_1!m_2!...m_N!$, and
\be
(x-q(t))^\mm = (x^1-q^1(t))^{m_1} (x^2-q^2(t))^{m_2} ...
 (x^N-q^N(t))^{m_N}.
\ee
Denote by $\mu$ a unit vector in the $\mu$:th direction, so that
$\mm+\mu = (m_1, \ab ...,m_\mu+1, \ab ..., \ab m_N)$, and let
\be
\fam(t) = \d_\mm\fa(q(t),t)
= \underbrace{\d_1 .. \d_1}_{m_1} .. 
\underbrace{\d_N .. \d_N}_{m_N} \fa(q(t),t)
\label{jetdef}
\ee
be the $|\mm|$:th order derivative of $\fa(x,t)$ evaluated on the
observer's trajectory $\qmu(t)$. Such objects transform as
\bes
[\Lxi, \fam(t)] &=& \d_\mm([\Lxi,\fa(q(t),t)]) 
+ [\Lxi,\qmu(t)]\dmu\d_\mm\fa(q(t),t) \nl
&\equiv& -\sumnm T^{\bt\nn}_{\al\mm}(\xi(q(t))) \fbn(t), \nl
{[}\J_X, \fam(t)] &=& \d_\mm([\J_X,\fa(q(t),t)]) 
\label{jet} \\
&\equiv& -\sumnm J^{\bt\nn}_{\al\mm}(X(q(t))) \fbn(t), \nl
{[}L_f, \fam(t)] &=& -f(t)\dot \phi_{\alpha,\mm}(t) 
- \la(\dot f(t)-if(t))\fam(t),
\eens
where
\bes
&&T^\mm_\nn(\xi) \equiv (T^{\al\mm}_{\bt\nn}(\xi)) \nl
&&={\nn\choose\mm} \d_{\nn-\mm+\nu}\xmu T^\nu_\mu 
 + {\nn\choose\mm-\mu}\d_{\nn-\mm+\mu}\xmu
  - \dlt^{\mm-\mu}_\nn \xmu,
\label{Tmn}\\
&&J^\mm_\nn(X) \equiv (J^{\al\mm}_{\bt\nn}(X))
= {\nn\choose\mm} \d_{\nn-\mm} X_a J^a,
\eens
and
\be
{\mm\choose\nn} = {\mm!\/\nn!(\mm-\nn)!} = 
{m_1\choose n_1}{m_2\choose n_2}...{m_N\choose n_N}.
\ee

We thus obtain a (non-linear) realization of $\vect(N)$ on the space of
trajectories in the space of tensor-valued
$p$-jets\footnote{$p$-jets are usually defined
as an equivalence class of functions: two functions are equivalent if all
derivatives up to order $p$, evaluated at $\qmu$, agree. However, each
class has a unique representative which is a polynomial of order at most
$p$, namely the Taylor expansion around $\qmu$, so we may canonically
identify jets with truncated Taylor series. Since $\qmu(t)$ depends on a
parameter $t$, we deal in fact with trajectories in jet space, but these
will also be called jets for brevity.}; 
denote this space by $J^p\QQ$. Note that $J^p\QQ$ is spanned by $\qmu(t)$
and $\{\fam(t)\}_{|\mm|\leq p}$ and thus not a $DGRO(N,\oj)$ module by
itself, because diffeomorphisms act non-linearly on $\qmu(t)$, as can
be seen in (\ref{DGRO}). However, the space $C(J^p\QQ)$ of functionals on
$J^p\QQ$ (local in $t$) {\em is} a module, because the action on a $p$-jet
can never produce a jet of order higher than $p$. The space
$C(q) \otimes_q J^p\QQ$, where only the trajectory itself appears 
non-linearly, is a submodule.

The crucial observation is that the jet space $J^p\QQ$
consists of finitely many functions of a single
variable $t$, which is precisely the situation where the normal ordering
prescription works. After normal ordering, denoted by double dots $:\ :$,
we obtain a Fock representation of the DGRO algebra:
\bes
\Lxi &=& \int dt\ \Big\{ \no{\xmu(q(t)) \pmu(t)} -
\sum_{|\nn|\leq|\mm|\leq p}
T^{\bt\nn}_{\al\mm}(\xi(q(t))) \no{ \fbn(t)\pam(t) } \Big\}, \nl
\J_X &=& -\int dt\ \Big\{ \sum_{|\nn|\leq|\mm|\leq p}
J^{\bt\nn}_{\al\mm}(\xi(q(t))) \no{ \fbn(t)\pam(t) } \Big\}, 
\label{Fock}\\
L_f &=& \int dt\ \Big\{ -f(t)\no{\dot \phi_{\alpha,\mm}(t)\pam(t)} 
- \la(\dot f(t)-if(t))\no{\fam(t)\pam(t)} \Big\},
\eens
where we have introduced canonical momenta
$\pmu(t) = \dd{\qmu(t)}$ and $\pam(t) = \dd{\fam(t)}$.
The field $\fa(x,t)$ can be either bosonic or fermionic but the
trajectory $\qmu(t)$ is of course always bosonic.

Normal ordering is defined with respect to frequency; any function of
$t \in S^1$ can be expanded in a Fourier series, e.g.
\be
\pmu(t) = \sum_{n=-\infty}^\infty \hat\pmu(n) \e^{-int} \equiv
\pmu^<(t) + \hat\pmu(0) + \pmu^>(t),
\ee
where $\pmu^<(t)$ ($\pmu^>(t)$) is the sum over negative (positive) 
frequency modes only. Then
\[
\no{\xmu(q(t)) \pmu(t)} 
\equiv \xmu(q(t)) \pmu^\leq(t) + \pmu^>(t) \xmu(q(t)),
\]
where the zero mode has been included in $\pmu^\leq(t)$.

It is clear that (\ref{Fock}) defines a Fock representation for every
$gl(N)$ irrep $\rep$ and every $\oj$ irrep $M$; denote this Fock space
by $J^p\FF$, which indicates that it also depends on the truncation 
order $p$. Namely, introduce a
Fock vacuum $\ket0$ which is annihilated by half of the oscillators,
e.g. $\qmu_<(t)$, $\pmu^\leq(t)$, $\fam^<(t)$ and $\pam_\leq(t)$. Then
$DGRO(N,\oj)$ acts on the space of functionals 
$C(\qmu_\geq,\pmu^>,\fam^\geq,\pam_>)$ of the remaining oscillators;
this is the Fock module.
Define numbers $k_0(\rep)$, $k_1(\rep)$, $k_2(\rep)$ and $y_M$ by
\bes
\tr{\rep} T^\mu_\nu &=& k_0(\rep) \dlt^\mu_\nu, \nl
\tr{\rep} T^\mu_\nu T^\si_\tau &=&
 k_1(\rep) \dlt^\mu_\tau \dlt^\si_\nu 
 + k_2(\rep) \dlt^\mu_\nu \dlt^\si_\tau, \\
\tr{M} J^aJ^b &=& y_M \dlt^{ab}.
\eens
The values of the abelian charges $c_1 - c_5$ (\ref{DGRO}) 
were calculated in \cite{Lar98}, Theorems 1 and 3, and in 
\cite{Lar01a}, Theorem 1:
\bes
c_1 &=& 1 - u\Np{} - x {N+p+1\choose N+2}, \nl
c_2 &=& -v\Np{} - 2w \Np{+1} - x {N+p\choose N+2}, \nl
c_3 &=& 1 + (1-2\la) ( w\Np{} + x\Np{+1}), 
\label{cs}\\
c_4 &=& 2N - x(1-6\la+6\la^2)\Np{}, \nl
c_5 &=& y\Np{}.
\eens
where 
\bes
u = \mp k_1(\rep)\, \dim\,M, &\qquad&
x = \mp \dim\,\rep\,\dim\,M, \nl
v = \mp k_2(\rep)\, \dim\,M, &\qquad&
y = \mp \dim\,\rep\,y_M, 
\label{numdef}\\
w = \mp k_0(\rep)\, \dim\,M, &&
\eens
and the sign factor depends on the Grassmann parity of $\fa$; the upper
sign holds for bosons and the lower for fermions, respectively.
The $p$-independent contributions to $c_1$, $c_3$ and $c_4$ come from
the trajectory $\qmu(t)$ itself.
We will henceforth set $\la = 0$.

\section{ Dynamics and the KT complex }

The modules constructed in the previous section show that a quantum
generalization of tensor calculus exists. However, this is not by
itself a theory of quantum gravity any more than tensor calculus 
determines general relativity. Somehow information about dynamics
must be included into the picture. A natural candidate is
found in the physics of gauge theories, as formulated cohomologically
in the anti-field formalism \cite{HT92}. 

The goal of classical physics is to find the {\em stationary surface}
$\Sigma$, i.e. the set of solutions to the Euler-Lagrange (EL) equations, 
viewed as a submanifold embedded in the space of all field configurations
$\QQ$.\footnote{ Note that a basis for this ``configuration space'' $\QQ$ 
is given by all fields in {\em spacetime}, not just in space.} Dually,
one wants to construct the function algebra $C(\Sigma) = C(\QQ)/\II$,
where $\II$ is the ideal generated by the EL equations. 
$C(\Sigma)$ will evidently carry a representation of the Noether 
symmetries of the action $S=\int \dNx\ \pounds(\phi)$. 

We can thus regard classical physics as the representation theory of its
Noether symmetries. The conventional next step would be to identify
gauge-equivalent configurations by passing to the orbit space, and thus
obtain a covariant description of phase space as the space of solutions to
the EL equation modulo gauges \cite{GR99}. However, this would be rather
uninteresting from an algebraic point of view, since the gauge symmetries
by construction act trivially on the orbit space. More importantly, as we
have seen above, normal ordering gives rise to non-trivial abelian
extensions, which could be interpreted as an {\em anomaly}; a putative
BRST charge would not remain nilpotent or even well defined. 
One can equivalently consider the full
stationary surface together with the action of the gauge symmetries.
Classically, and in the absense of anomalies, the two formulations are
equivalent; the
passage to the orbit space can always be performed if so desired. The
abelian extension becomes harmless\footnote{
The extension would be very harmful if representations of the
extended algebra were missing.}
from this point of view; the algebra
merely acquires its full quantum form but the representation theory
remains well defined. Indeed, all interesting representations of the the
diffeomorphism algebra, at least in one dimension, have non-zero
extensions. This issue is further discussed in the conclusion.

The Lagrangian $\pounds(\phi)$ is 
a local functional of $\phi$, i.e. a function of $\fa(x)$ and its 
derivatives $\dmu\fa(x)$, $\dmu\dnu\fa(x)$, etc., up to some finite
order, all evaluated at the same point $x$. In practice, the Lagrangian
only depends on first-order derivatives. The EL equations,
\be
\Ea(x) \equiv {\dlt S\/\dlt\fa(x)} 
= {\d\pounds\/\d\fa}(x) - {\dmu}{\d\pounds\/\dmu\fa}(x) = 0,
\label{EL}
\ee
generate an ideal $\II\subset C(\QQ)$, and the factor space 
$C(\Sigma) = C(\QQ)/\II$ is still a $\dmap$ module due to the invariance
assuption. This factor space is
most conveniently described as a resolution of a certain Koszul-Tate (KT)
complex. For each field $\fa(x)$, introduce an antifield $\fsa(x)$ 
transforming as the corresponding EL equation $\Ea(x)$. We then consider
the extended configuration space $\QQ^*$ as the span of $\fa(x)$ and 
$\fsa(x)$. Now consider the space of local functionals on $\QQ^*$:
$C(\QQ^*) = C(\phi,\fs)$.
If $\phi$ is bosonic ($C(\phi)$ consists of symmetric functionals), then
$\fs$ is fermionic ($C(\fs)$ consists of anti-symmetric functionals), and 
vice versa. 

Define the {\em anti-field number} by $\afn\fa = 0$, $\afn\fsa = 1$.
$C(\QQ^*)$ can be decomposed into subspaces $C^g(\QQ^*)$ of fixed antifield
number $g$:
\be
C(\QQ^*) = \oplus_{g=0}^\infty C(\phi)\otimes C^g(\fs)
\equiv \oplus_{g=0}^\infty C^g(\QQ^*).
\ee
The KT complex takes the form
\be
0 \larroww \dlt C^0(\QQ^*) \larroww \dlt C^1(\QQ^*) \larroww \dlt 
C^2(\QQ^*)\larroww \dlt \ldots
\label{complex1}
\ee
where the KT differential $\dlt$ is defined by
\be
\dlt\fa(x) = 0, \qquad
\dlt\fsa(x) = \Ea(x).
\ee
By a standard argument \cite{HT92}, the cohomology groups $H^g(\dlt) = 0$
unless $g=0$, and $H^0(\dlt) =C(\QQ)/\II = C(\Sigma)$. We have thus
obtained a resolution of the space of functionals on the stationary
surface, as desired.

Introduce canonical momenta $\pa(x)=\dd{\fa(x)}$ and 
$\psa(x)=\dd{\fsa(x)}$, with antifield numbers $\afn\pa = 0$,
$\afn\psa = -1$.
The KT differential can then be written as a bracket: 
$\dlt F = [Q, F]$, where
\be
Q = \int \dNx\ \Ea(x)\psa(x).
\label{Q0}
\ee
Let $\PP$ be the space spanned by $\fa(x)$ and $\pa(x)$, and let 
$\PP^*$ be the span of $\fa(x)$, $\fsa(x)$, $\pa(x)$ and $\psa(x)$.
The expression (\ref{Q0}) defines a differential, also denoted by $Q$, 
which acts on the space $C(\PP^*)$ of local functionals on $\PP^*$. 
Note that
$C(\PP^*)$ is a non-commutative ring, which can be thought of as the
algebra of differential operators on $\QQ^*$. The decomposition into
subspaces of fixed antifield number now extends indefinitely in both 
directions:
\be
C(\PP^*) = \oplus_{g=-\infty}^\infty C^g(\PP^*).
\ee
Accordingly, we obtain the two-sided complex
\be
\ldots \larroww Q C^{-1}(\PP^*) \larroww Q C^0(\PP^*)
\larroww Q C^1(\PP^*) \larroww Q \ldots
\label{complex3}
\ee
The cohomology group $H^0(Q)$ can be thought of as the space of 
differential operators on the stationary surface $\Sigma$. However, I
do not know if (\ref{complex3}) is a resolution, i.e. if the other
cohomology groups vanish.

There is a problem: the EL equations may be dependent, i.e. there 
may be relations of the form
\be
r^a(x) = r^a_\al(x)\Ea(x) \equiv 0,
\label{ra}
\ee
where $r^a_\al(x)$ is some functional of $\fa(x)$.
Then $H^1(Q) \neq 0$, because $r^a_\al(x)\fsa(x)$ is KT closed:
$[Q, r^a_\al(x)\fsa(x)] = 0$. The standard way to kill this unwanted 
cohomology is to introduce a second-order antifield $b^a(x)$. Let
$[Q, b^a(x)] = r^a_\al(x)\fsa(x)$, which makes the latter expression
exact and thus makes it vanish in cohomology. To obtain the explicit 
expression for $Q$, introduce the second-order antifield momentum
$c_a(x) = \dd{b^a(x)}$. The full KT differential is
now
\be
Q = \int \dNx\ (\Ea(x)\psa(x) + r^a_\al(x)\fsa(x)c_a(x) ).
\label{Q}
\ee
There can in principle be relations also among the $r^a_\al(x)$ of the
form $Z^A(x) = Z^A_a(x) r^a_\al(x) \equiv 0$. If so, it is necessary to
introduce higher-order antifields to eliminate the unwanted cohomology.
However, we will assume that the gauge symmetries are irreducible, i.e.
that no non-trivial higher-order relations exist, since this is the case
in all experimentally established theories of physics.

The situation is summarized in the following table:
\be
\barr{|c|c|c|l|}
\hline
g &  \hbox{Field} & \hbox{Momentum} & \hbox{Ideal} \\
\hline
0 & \fa(x) &\pa(x) & - \\
1 & \fsa(x) &\psa(x) & \Ea(x)\approx 0 \\
2 & b^a(x) &c_a(x) & r^a_\al(x)\fsa(x) \approx 0 \\
\hline
\earr
\ee

As a preparation for normal ordering,
we must now add the variable $t$, i.e. replace 
$\fa(x) \to \fa(x,t)$. 
The EL equations (\ref{EL}) now read $\Ea(x,t)=0$, and
the KT charge (\ref{Q}) is replaced by
\be
Q = \int \dxdt\ (\Ea(x,t)\psa(x,t) + r^a_\al(x,t)\fsa(x,t)c_a(x,t) ).
\ee
Since the space of functionals over $\fa(x,t)$ is larger than $C(\QQ)$, 
we must
factor out a larger ideal to obtain a resolution of the same space
$C(\Sigma)$. It is easy to see that the necessary additional requirement 
is $\d_t\fa(x,t)\approx0$; to implement this 
constraint in cohomology, we introduce the antifield $\wfa(x,t)$ with
canonical momentum $\wpa(x,t)$. Since $\Ea(x,t)$ depends on $\fa(x,t)$
only, we now have $\d_t\Ea(x,t)\equiv 0$,
which generates unwanted cohomology.
This is eliminated by introducing a second-order antifield $\wfsa(x,t)$.
Finally, the other second-order antifield $b^a(x,t)$, associated with
the gauge symmetry, is now reducible. Correct this by introducing a 
third-order antifield $\wba(x,t)$. 
The situation is summarized in the following table:
\be
\barr{|c|c|c|l|}
\hline
g & \hbox{Field} & \hbox{Momentum} & \hbox{Ideal} \\
\hline
0 & \fa(x,t)& \pa(x,t) & - \\
1 & \fsa(x,t) & \psa(x,t) & \Ea(x,t)\approx 0 \\
1 & \wfa(x,t)& \wpa(x,t) & \d_t\fa(x,t) \approx 0 \\
2 & b^a(x,t)& c_a(x,t) & r^a_\al(x,t)\fsa(x,t) \approx 0 \\
2 & \wfsa(x,t)& \wpsa(x,t) & \d_t\fsa(x,t) \approx 0 \\
3 & \wba(x,t)& \wca(x,t) & \d_t b^a(x,t) \approx 0 \\
\hline
\earr
\ee

\section{ KT complex in jet space and quantization }

In order to construct the jet space version of the KT complex, we
expand not only the fields but also the EL equations and the anti-fields 
in multi-dimensional Taylor series. Set 
$\Eam(t) = \d_\mm\Ea(q(t),t)$ and $\fsam(t) = \ab\d_\mm\fsa(q(t),t)$.
What must be noted is that we can only define $\Eam(t)$ for 
$|\mm|\leq p-\ord_\al$, where $\ord_\al$ is the order of the EL equation 
$\Ea(x)$. This is because $\Eam(t)$ is a function of $\fan(t)$ for all
$|\nn| \leq |\mm|+\ord_\al$, and $\fan(t)$ is undefined for $|\nn|>p$.
Similarly, the relations (\ref{ra}) and the corresponding
second-order anti-fields $b^a(x)$ give rise to the jets
$\ram(t) = \d_\mm(r^a_\al(q(t),t) \fsa(q(t),t))$ and 
$\bam(t) = \d_\mm b^a(q(t),t)$, respectively. 
If the relations $r^a(x)$ are of order $\ordg_a$ in the derivatives, 
$\ram(t)$ and $\bam(t)$ is only defined for $|\mm|\leq p-\ordg_a$.

The ideals of type $\d_t\fa(x,t) \approx 0$ translate into:
\bes
D_t\fam(t) &\equiv& \dot \phi_{\alpha,\mm}(t) - \dot q^\mu(t)\phi_{\al,\mm+\mu}(t)
\approx 0, \nl
D_t\fsam(t) &\equiv& \dot \phi^{*\alpha}_{,\mm}(t) - \dot q^\mu(t)\fsa_{,\mm+\mu}(t)
\approx 0, 
\label{D_t}\\
D_t\bam(t) &\equiv& \dot b^a_{,\mm}(t) - \dot q^\mu(t)b^a_{,\mm+\mu}(t)
\approx 0.
\eens
These conditions are implemented in cohomology by the introduction of
further (second and third order) anti-fields $\wfam(t)$, $\wfsam(t)$
and $\wbam(t)$. The conditions in (\ref{D_t}), and hence the barred
antifields, are only defined for one order less than the corresponding
unbarred antifield, since $|\mm+\mu| = |\mm|+1$.

Add dual coordinates (jet momenta) $\pmu(t) = \dd{\qmu(t)}$, 
$\pam(t) = \dd{\fam(t)}$, $\psam(t) = \dd{\fsam(t)}$, 
$\cam(t) = \dd{\bam(t)}$, $\wpam(t) = \dd{\wfam(t)}$,
$\wpsam(t) = \dd{\wfsam(t)}$ and $\wcam(t) = \dd{\wbam(t)}$.
Denote the space spanned by all jets and jet momenta by $J^p\PP^*$ and
the ring of local functionals on $J^p\PP^*$ by $C(J^p\PP^*)$; 
it may be considered as the differential operators on $J^p\QQ^*$.
The full KT differential acting on $C(J^p\PP^*)$ becomes
\bes
Q &=& \int dt\ \Big(\summ{p-\ord_\al}\Eam(t)\psam(t) 
+ \summ{p-\ordg_a}\ram(t)\cam(t) \nl
&&+ \summ{p-1}D_t\fam(t)\wpam(t) 
+ \summ{p-\ord_\al-1}D_t\fsam(t)\wpsam(t) \nl
&&+ \summ{p-\ordg_a-1}D_t\bam(x,t)\wcam(x,t) \Big).
\label{Qjet}
\ees
The situation is summarized in the following table:
\be
\barr{|c|c|c|c|c|}
\hline
g & \hbox{Jet} & \hbox{Momentum} & \hbox{Order} &\hbox{Ideal} \\
\hline
0 & \fam(t)& \pam(t) & p & - \\
1 & \fsam(t) & \psam(t) & p-\ord_\al & \Eam(t)\approx 0 \\
1 & \wfam(t)& \wpam(t) & p-1 & D_t\fam(t) \approx 0 \\
2 & \bam(t)& \cam(t) & p-\ordg_a &\ram(t) \approx 0 \\
2 & \wfsam(t)& \wpsam(t) & p-\ord_\al-1 &D_t\fsam(t) \approx 0 \\
3 & \wbam(t)& \wcam(t) & p-\ordg_a-1 &D_t\bam(t) \approx 0 \\
\hline
\earr
\ee

We are now ready for the quantization step. The spaces $C^g(\QQ^*)$ in 
the complex (\ref{complex1}) are field spaces to which the Fock 
construction in section \ref{sec:Fock} applies.
Thus, we simply replace $C^g(\PP^*) \to J^p\FF^g$ in (\ref{complex3}) 
and obtain the complex
\be
\ldots \larroww Q J^p\FF^{-2} \larroww Q J^p\FF^{-1} 
\larroww Q J^p\FF^0
\larroww Q J^p\FF^1 \larroww Q J^p\FF^2 \larroww Q \ldots
\ee
A crucial observation is that the KT charge $Q$ is a bilinear combination
of graded-commutative terms. Hence it is not affected
by normal ordering, and the cohomology groups of this complex are
well defined $DGRO(N,\oj)$ modules since $Q$ commutes with the module
action. It would not be possible to 
construct a similar complex with a BRST charge, because normal ordering
would then ruin nilpotency.

\section {Finiteness condition}

The modules obtained in this fashion are well 
defined for all finite values of the jet order $p$, but in order to
have a field theory interpretation, it must be possible
to reconstruct the original field by means of the Taylor series
(\ref{Taylor}), i.e. to take the limit $p\to\infty$. A necessary
condition for taking this limit is that the abelian charges have a
finite limit. 
Taken at face value, the prospects for succeeding
appear bleak. When $p$ is large, ${m+p\choose n} \approx p^n/n!$, so 
the abelian charges (\ref{cs}) diverge; the worst case is 
$c_1 \approx c_2 \approx p^{N+2}/(N+2)!$, which diverges in all dimensions
$N > -2$. In \cite{Lar01a} a way out of this problem was devised: consider 
a more general realization by taking the direct sum of operators 
corresponding to different values of the jet order $p$. Take
the sum of $r+1$ terms like those in (\ref{Fock}), with $p$ replaced by
$p$, $p-1$, ..., $p-r$, respectively, and with $\rep$ and $M$ replaced by
$\repi$ and $\Mi$ in the $p-i$ term. 

Such a sum of contributions arises naturally from the KT complex, because
the antifields are only defined up to an order smaller than $p$ (e.g.
$p-\ord_\al$ or $p-\ordg_a$). 
Denote the numbers $u,v,w,x,y$ in the modules $\repi$ and $\Mi$,
defined as in (\ref{numdef}), by $u_i,v_i,w_i,x_i,y_i$, 
respectively. Of course, there is 
only one contribution from the observer's trajectory. 
Then it was shown in \cite{Lar01a}, Theorem 3, that
\bes
c_1 = -U\Npr, &\qquad&
c_2 = -V\Npr, \nl
c_3 = W\Npr, &\qquad&
c_4 = -X\Npr,
\label{finc} \\
c_5 = Y\Npr, 
\eens
where $u_0=U$, $v_0=V$, $w_0=W$, $x_0=X$ and $y_0=Y$, provided that the 
following conditions hold:
\bes
i&\qquad&u_i + \ritwo X = \mri U, \nl
ii&\qquad&v_i - 2\rione W - \ritwo X = \mri V, \nl
iii&\qquad&w_i - \rione X = \mri W, 
\label{conds}\\
iv&\qquad&x_i = \mri X, \nl
v&\qquad&y_i = \mri Y.
\eens
The contributions from the observer's trajectory have also been eliminated
by antifields coming from the geodesic equation \cite{Lar02}; this is
not important in the sequel because these contributions were finite 
anyway.

\section{Solutions to the finiteness conditions}
Let us now consider the solutions to (\ref{conds}) for the numbers
$x_i$, which can be interpreted as the number of fields and anti-fields.
First assume
that the field $\fam(t)$ is fermionic with $x_F$ components, which gives
$x_0=x_F$. We may assume, by the spin-statistics theorem, that the 
EL equations are first order, so the bosonic antifields $\fsam(t)$ 
contribute $-x_F$ to $x_1$. The barred antifields $\wfam(t)$ are also
defined up to order $p-1$, and so give $x_1=-x_F$, and the barred 
second-order antifields $\wfsam(t)$ give $x_2=x_F$. Further assume that
the fermionic EL equations have $x_S$ gauge symmetries, i.e. the 
second-order antifields $\bam(t)$ give $x_2=x_S$. In established
theories, $x_S=0$, but we will need a non-zero value for $x_S$.
Finally, the corresponding barred antifields give $x_3 = -x_S$.

For bosons the situation is analogous, with two exceptions: all signs
are reversed, and the EL equations are assumed to be second order.
Hence $\fsam(t)$ yields $x_2=x_B$ and the gauge antifields $\bam(t)$ 
give $x_3=-x_G$. Accordingly, the barred antifields are one order higher.

The situation is summarized in the following tables, where the upper half
is valid if the original field is fermionic and the lower half if it is
bosonic:
\bes
\barr{|c|c|c|l|}						       
\hline
g & \hbox{Jet} & \hbox{Order} &x \\
\hline
0 & \fam(t) & p & x_F \\
1 & \wfam(t) & p-1 & -x_F \\
1 & \fsam(t) & p-1 & -x_F \\
2 & \wfsam(t) & p-2 & x_F \\
2 & \bam(t) & p-2 & x_S \\
3 & \wbam(t) & p-3 & -x_S \\
\hline
\hline
0 & \fam(t) & p & -x_B \\
1 & \wfam(t) & p-1 & x_B \\
1 & \fsam(t) & p-2 & x_B \\
2 & \wfsam(t) & p-3 & -x_B \\
2 & \bam(t) & p-3 & -x_G \\
3 & \wbam(t) & p-4 & x_G \\
\hline
\earr
\label{tab}
\ees
If we add all contributions of the same order, we see that relation 
$iv$ in (\ref{conds}) can only be satisfied provided that
\bes
p: &\quad& x_F - x_B = X \nl
p-1: && -2x_F+x_B = -rX, \nl
p-2: && x_B + x_F + x_S = {r\choose2}X, \nl
p-3: && -x_B-x_S-x_G = -{r\choose3}X, 
\label{rcond}\\
p-4: && x_G = {r\choose4}X, \nl
p-5: && 0 = -{r\choose5}X, ...
\eens
The last equation holds only if $r\leq4$ (or trivially if $X=0$). 
On the other hand, if we demand that there is at least one bosonic
gauge condition, the $p-4$ equation yields $r\geq4$. Such a demand is
natural, because both the Maxwell/Yang-Mills and the Einstein equations
have this property. Therefore, we are unambigiously guided to consider
$r=4$ (and thus $N=4$). The specialization of (\ref{rcond}) to four
dimensions reads
\bes
p: &\quad& x_F - x_B = X \nl
p-1: && -2x_F+x_B = -4X, \nl
p-2: && x_B + x_F + x_S = 6X, \\
p-3: && -x_B-x_S-x_G = -4X, \nl
p-4: && x_G = X.
\eens
Clearly, the unique solution to these equations is
\be
x_F = 3X, \qquad x_B = 2X, \qquad x_S = X, \qquad x_G = X.
\label{xsol}
\ee
The solutions to the remaining equations in (\ref{conds}) are found by
analogous reasoning. The result is
\bes
\barr{llllll}
u_B = 2U  &\qquad& v_B = 2V+2W \\
u_F = 3U  && v_F = 3V+2W \\
u_S = U-X && v_S = V+2W+X \\
u_G = U-X && v_G = V+2W+X \\
\\
w_B = 2W+X && y_B = 2Y \\
w_F = 3W+X && y_F = 3Y \\
w_S = W+X  && y_S = Y \\
w_G = W+X  && y_G = Y \\
\\
\earr
\ees
This result expresses the twenty parameters $x_B-w_G$
in terms of the five parameters $X$, $Y$, $U$, $V$, $W$.
For this particular choice of parameters, the abelian charges in
(\ref{finc}) are given by 
\be
c_1 = -U, \qquad c_2 = -V, \qquad c_3 = W, \qquad c_4 = -X,
\qquad c_5 = Y,
\ee
independent of $p$. Hence there is no manifest obstruction to the
limit $p\to\infty$.\footnote{After this work was completed, I realized 
that the negative signs of $c_1$, $c_2$ and $c_4$ imply problems with
unitarity. Finiteness seems to be a more pressing problem, however.}

\section{Comparison with known physics}
All experimentally known physics is well described by quantum theory, 
gravity, and the standard model in four dimensions.
We have already seen that quantum general covariance more or less 
dictates that spacetime has $N=4$ dimensions (\ref{rcond}). It is therefore
interesting to investigate to what extent the particle content 
matches (\ref{xsol}); recall that $x = \tr{} 1$ equals the number of field
components.

The bosonic content of the theory is given by the following table. Standard
notation for the fields is used, and one must remember that it is the 
na\"\i ve number
of components that enters the equation, not the gauge-invariant physical
content. E.g., the photon is described by the four components $A_\mu$ 
rather than the two physical transverse components.
Also, the gauge algebra $\ssu$ has $8+3+1 = 12$ generators.
\bes
\barr{|c|c|c|c|}
\hline
\hbox{Field} & \hbox{Name}& \hbox{EL equation} & x_B  \\
\hline
A^a_\mu & \hbox{Gauge bosons}
&D_\nu F^{a\mu\nu} = j^{a\mu} &12\times4 = 48  \\
g_{\mu\nu} & \hbox{Metric}
&G^{\mu\nu} = {1\/8\pi}T^{\mu\nu} & 10 \\
H & \hbox{Higgs field}
& g^{\mu\nu}\dmu\dnu H = V(H) & 2\\
\hline
\earr
\nle
\barr{|c|c|}
\hline
\hbox{Gauge condition} & x_G  \\
\hline
D_\mu D_\nu F^{a\mu\nu} = 0 & 12\times1 = 12 \\
\dnu G^{\mu\nu} = 0 & 4 \\
\hline
\earr
\eens
The total number of bosons in the theory is thus $x_B = 48+10+2 = 60$,
which implies $X=30$ by (\ref{xsol}). The number of gauge conditions is
$x_G = 16$, which implies $X=16$. There is certainly a discrepancy here.

The fermionic content in the first generation is given by
\bes
\barr{|c|c|c|c|}
\hline
\hbox{Field} & \hbox{Name}& \hbox{EL equation} & x_F  \\
\hline
u & \hbox{Up quark} & \Dslash u = ... & 2\times3 = 6\\
d & \hbox{Down quark} & \Dslash d = ... & 2\times3 = 6\\
e & \hbox{Electron} & \Dslash e = ... & 2\\
\nu_L & \hbox{Left-handed neutrino} & \Dslash \nu_L = ... & 1\\
\hline
\earr
\ees
The number of fermions in the first generation is thus 
$x_F = 6+6+2+1 = 15$. Counting all three generations and anti-particles,
we find that the total number of fermions is $x_F = 2\times3\times15 = 90$,
which implies $X=30$. There are no fermionic gauge conditions, so $x_S=0$,
which implies $X=0$.

It is clear that the predictions for $X$ ($30,16,30,0$) are not mutually
consistent. However, to cancel the leading terms, of order $p$ and $p-1$,
it is only necessary that $2x_F = 3x_B$, which is indeed the case in known
physics. It is therefore tempting to speculate that known physics is a
first approximation of a more elegant theory, which has the same field
content but more gauge conditions, including fermionic ones. An attractive
possibility, suggested by Kac \cite{Kac99}, would be to replace the
standard model symmetries by one of the recently discovered exceptional
Lie superalgebras, whose irreps are in 1-1 correspondence with
$\ssu$ irreps.

It is important to check that the results remain the same if the same 
physical situation is described with a different, but equivalent, set of
fields. Typically, such spurious degrees of freedom have algebraic
EL equations. Denote the original (bosonic, say) $x_B$ fields by $\fam(t)$
and let $\psi_{i,\mm}(t)$ be $x_A$ spurious fields, defined for 
$|\mm|\leq p$. The contribution to $x_0$ from the bosonic fields is thus
$-x_B - x_A$. There are also $x_A$ new EL equations $E^i_{,\mm}(t)$,
defined for $|\mm|\leq p$ because they are algebraic; $E^i_{,\mm}(t)$
contains $\psi_{j,\nn}(t)$ for all $|\nn|\leq|\mm|$, but not of higher
order. The corresponding anti-fields $\psi^{*i}_{,\mm}(t)$ add $x_A$ to
$x_0$. The total result is $x_0 = -x_B-x_A+x_A = -x_B$, as before.

An example is given by the gravitational field in vielbein formalism.
Instead of the ten components of the metric $g_{\mu\nu} = g_{\nu\mu}$
we have the sixteen vielbein components $e^i_\mu$. However, the 
requirement that the 
metric $g_{\mu\nu} = e^i_\mu e_{i\nu}$ be symmetric gives rise to 
six algebraic conditions, so the contribution to $x_0$ is still ten.

\section{Conclusion}

There are two key lessons to be learnt from twentieth century physics:
\begin{itemize}
\item
General relativity teaches us the importance of diffeomorphism invariance.
Physics is fully relational; there is no background stage over
which physics takes place, but geometry itself participates actively
in the dynamics. Note that this is very different from mere coordinate
invariance, because there is no compensating background metric.
\item
Quantum theory teaches us the importance of projective lowest-energy
representations; the passage from Poisson brackets to commutators
makes normal ordering necessary, and the brackets typically acquire
quantum corrections.
\end{itemize}
The successful construction of a quantum theory of gravity will probably
combine these two insights. It seems obvious that the correct way to
combine diffeomorphism invariance and projective representations is to
consider projective representations of the diffeomorphism group, which
on the Lie algebra level gives rise to the DGRO algebra.

A common objection is that the presence of an extension makes
diffeomorphism symmetry anomalous. Although anomaly cancellation is
certainly a valuable mechanism which is experimentally confirmed in the
standard model, it is not so natural from an algebraist's point of view;
in particular, all mathematically interesting (= non-trivial, irreducible,
unitary) representations of the Virasoro algebra have a positive value of
the central charge, something that is also necessary for locality, 
i.e. decaying correlation functions.
An intriguing recent observation is that
post-Newtonian corrections seem to violate general covariance
\cite{Kaz03}; this is possibly related because an extension is 
the simplest way to relax diffeomorphism symmetry in a mathematically 
consistent way.
Moreover, the Schwinger terms arising in the standard model are quite
different from the multi-dimensional Virasoro algebra. 
They give rise to Mickelsson-Faddeev algebras, which are known
not to possess fully quantum representations \cite{Pic89}. As shown in
this paper, not only does the DGRO algebra possess quantum
representations, but one can associate a family of such representations
(labelled by the jet order $p$) to every general-covariant dynamical
system, which can probably be viewed as a kind of quantization.

\end{document}